\documentclass[%
 reprint,
 amsmath,amssymb,
 aps,
]{revtex4-2}
\usepackage{tikz}
\newcommand{\mycircle}[1]{%
\tikz[baseline]\node[draw,shape=circle,anchor=base,inner sep=0.4ex] {#1} ;}
\usepackage{graphicx}
\usepackage{dcolumn}
\usepackage{bm}

\begin{document}


\title{On the Formation of Super-Alfv\'enic Flows Downstream of Collisionless Shocks} 
\author{Adnane Osmane}
 \email{adnane.osmane@helsinki.fi}
\affiliation{University of Helsinki, Department of Physics, Helsinki, Finland
}%
\author{Savvas Raptis}
 \email{savvas.raptis@jhuapl.edu}
\affiliation{Johns Hopkins University Applied Physics Laboratory, Laurel, MD, USA
}%

\date{\today}

\begin{abstract}
\noindent Super-Alfv\'enic jets, with kinetic energy densities significantly exceeding that of the solar wind, are commonly generated downstream of Earth’s bow shock under both high and low beta plasma conditions. In this study, we present theoretical evidence that these enhanced kinetic energy flows are driven by firehose-unstable fluctuations and compressive heating within collisionless plasma environments. Using a fluid formalism that incorporates pressure anisotropy, we estimate that the downstream flow of a collisionless plasma shock can be accelerated by a factor of 2 to 4 following the compression and saturation of firehose instability. By analyzing quasi-parallel magnetosheath jets observed \textit{in situ} by the Magnetospheric Multiscale (MMS) mission, we find that approximately 11\% of plasma measurements within these jets exhibit firehose-unstable fluctuations. Our findings offer an explanation for the distinctive generation of fast downstream flows in both low ($\beta<1$) and high ($\beta>1$) beta plasmas, and provide new evidence that kinetic processes are crucial for accurately describing the formation and evolution of magnetosheath jets.
\end{abstract}

\maketitle

\section{Introduction}\label{sec:level1}

Shocks are ubiquitous in astrophysical plasma environments and serve as some of the most efficient sites for particle acceleration in the universe. These shocks are generated whenever supersonic and super-Alfvénic stellar flows collide with obstacles such as planetary or interstellar magnetic fields. During such interactions, the kinetic energy density of the stellar wind is converted into thermal and magnetic energy, resulting in a compressed and heated plasma, and an increase in entropy \cite{Kulsrud}. However, this conventional understanding of astrophysical shocks has been challenged in recent decades by the discovery of a puzzling phenomenon: magnetosheath jets \cite{nvemevcek1998transient, Savin08, Hietala09, Archer12}. These jets, first observed \textit{in situ} downstream of Earth’s bow shock \footnote{Recent studies also shows that jets are generated in the magnetosheath of other planets such as Mars  and Jupiter \cite{gunell2023magnetosheath,zhou2024magnetosheath} and downstream of interplanetary shocks \cite{hietala2024candidates_interplanetary}. Since jets are  commonly observed downstream of quasi-parallel supercritical shocks they should also be present in a wide range of high Mach number astrophysical shocks of other planetary environments.}, have been shown to exhibit flow velocities that are comparable to those of the upstream flow \footnote{The magnetospheric literature describes jets in terms of dynamic pressure $\rho u^2$, where $\rho$ is the mass density and $u$ is the flow velocity \cite{plaschke2018jets}. In this communication we prefer to describe jets in terms of kinetic energy density  $\rho u^2/2$ since it is the quantity that enters in the conservation of energy equation for the macroscopic plasma model of \cite{CGL} that we use in Section 2.}. This paradox raises a fundamental question in fundamental plasma physics: How can a supersonic and super-Alfvénic flow cross a shock, decelerate, and heat the plasma, yet emerge with increased kinetic energy density? Resolving this apparent contradiction is crucial for advancing our understanding of shock dynamics in astrophysical plasmas, with significant implications for particle acceleration processes \cite{perri2022recent}, particularly since the efficiency of diffusive shock acceleration is dependent on the difference between upstream and downstream flow velocities \cite{fermi1949origin, caprioli2014simulations, park2015simultaneous}.

Since the discovery of magnetosheath jets, several mechanisms have been suggested to explain their generation. \citet{Hietala13} suggested a mechanism in which a localised ripple in the bow shock could lead to the refraction and penetration of the plasma without dissipation, such that the flow downstream of the shock is effectively comparable to the flow upstream.  \citet{Archer12} suggested that jets are caused by solar wind pressure pulses or rotational discontinuities that can enhance kinetic energy density downstream of the shock. \citet{karlsson2015origin} associated the formation of jet with transient coherent structures observed upstream of the Earth's bow shock known as short large-amplitude magnetic structures (SLAMS). SLAMS are steepened foreshock fluctuations with thermal and magnetic field energy sometimes several multiples of the kinetic energy density in the solar wind. More recently,   \citet{raptis2022downstream} have shown direct observations of downstream super-magnetosonic jets generated directly from the evolution of upstream fluctuations and the shock reformation cycle, thus suggesting a mechanism that relies on kinetic scale instabilities. 

Statistical studies of spacecraft data by \citet{Plaschke13} and \citet{raptis2020classifying} found evidence for both the bow shock ripple mechanism suggested by \citet{Hietala13} and the SLAMS mechanism of \citet{karlsson2015origin}. However, from an observational perspective the problem with trying to prove these theories based on observations is that the spacecraft are seldom, if ever, in a suitable array such that both the production mechanism and the resulting jet can simultaneously be identified unambiguously. This task is now made much more difficult by the recent recognition that the majority of jets are much smaller in scale than previously reported \cite{plaschke2020scale}, and that jets are more likely to form at kinetic scales \cite{raptis2022magnetosheath}. 

Moreover, from a theoretical perspective, whether one aims to quantify the contribution to jets' formation from the ripple mechanism of \citet{Hietala13} or from the mechanisms suggested by \citet{karlsson2015origin} or \citet{raptis2022downstream}, all rely on the presence of kinetic structures and instabilities, such as SLAMS, either to deform the bow shock surface, or to release their thermal and magnetic field energy into kinetic energy. Every suggested mechanism therefore relies on the presence of kinetic scale structures, but the energetic contribution of kinetic processes in the generation of jets has yet to be quantified \textit{analytically} \footnote{Analytical studies of magnetosheath jets that incorporate kinetic processes are currently missing but it should be pointed out that kinetic-hybrid simulations have been conducted. For instance see \citet{palmroth2021magnetosheath, suni2021connection, omelchenko20213d, preisser2020magnetosheath} and references therein. Numerical studies are necessary to characterise magnetosheath jets but need to be complemented by analytical studies to determine if the kinetic structures observed in simulations and \textit{in situ} hold sufficient energy to explain jets formation.}. Thus, the origin of magnetosheath jets remains to this day an open question in parts because the energy sources at kinetic and fluid scales have not been determined and quantified \footnote{See \citet{plaschke2018jets} for a more recent summary of numerical and observational results}. 

In this communication, we identify for the first time the mechanisms at the interface of fluid and kinetic processes that can generate magnetosheath jets. By employing the Chew-Goldberger-Low (CGL) fluid formalism, which integrates kinetic effects such as pressure anisotropy \cite{CGL}, we uncover multiple contributing mechanisms to the enhanced kinetic energy density at macroscopic scales. Our theoretical approach enables us to precisely isolate the influence of kinetic processes and determine the plasma conditions under which alternative mechanisms may dominate. 

The article is organised as follow. In Section 2, we present the theoretical model based on the CGL equations and focus on two sources for jets generation: (1) firehose unstable plasma and (2) compressive and rarefied plasmas for low and higher beta plasma environment, respectively. In Section 3, we provide initial observational verification to our theoretical analysis by using measurements from Magnetospheric Multiscale (MMS) mission \cite{burch2016magnetospheric} downstream of Earth's bow shock to determine if the plasma inside of jets is firehose unstable. In Section 4, we summarise our results by arguing that any study of jets' formation would require the incorporation of kinetic processes.

\section{Theoretical Methodology}
\subsection{Macroscopic dynamics due to kinetic processes}
We consider the simplest set of equations for describing a macroscopic plasma that incorporates kinetic effects. In this formalism, the plasma is fully ionized, and the pressure tensor is gyrotropic. However, pressure anisotropy—where the pressure parallel and perpendicular to the local magnetic field lines differ—can be generated, sustained, and dynamically tracked. This approximation is valid for motions occurring on spatial and temporal scales significantly larger than those associated with ion gyromotion. For jets much larger than Larmor scales, this low-frequency, long-wavelength limit provides a solid foundation for analytically estimating the impact of kinetic processes on macroscopic quantities such as density and mean flow. It leads to the following macroscopic equations for the magnetic field and the first three moments of the plasma distribution function \cite{CGL, kulsrud1983mhd, schekochihin2010magnetofluid, squire2017amplitude},  
\begin{equation}
\frac{\partial\rho}{\partial t}+\nabla\cdot (\rho \mathbf{u})=0, \nonumber
\end{equation}
\begin{eqnarray}
\rho\left(\frac{\partial\mathbf{u}}{\partial t}+\mathbf{u}\cdot
\mathbf{\nabla}\mathbf{u}\right)&=&-\nabla
\left(p_\perp+\frac{B^2}{8\pi}\right) \nonumber  \\&+&\nabla\cdot\left[\mathbf{\mathbf{b}\mathbf{b}}
\left(\Delta p+\frac{B^2}{4\pi}\right)\right], \nonumber
\end{eqnarray}
\begin{equation}
\label{Initial_eq}
\frac{\partial \mathbf{B}}{\partial t}=\nabla \times (\mathbf{u}\times \mathbf{B}), 
\end{equation}
\begin{eqnarray}
\frac{\partial p_\perp}{\partial t}+\nabla\cdot(p_\perp\mathbf{u})+p_\perp \nabla \cdot \mathbf{u}&=&p_\perp \mathbf{\mathbf{b}\mathbf{b}}\ \colon\nabla\mathbf{u}-2\nu_c\Delta p \nonumber \\
&-&\nabla \cdot (q_\perp \mathbf{\mathbf{b}})-q_\perp \nabla\cdot\mathbf{\mathbf{b}}, \nonumber
\end{eqnarray}
\begin{equation}
\frac{\partial p_\parallel}{\partial t}+\nabla\cdot(p_\parallel\mathbf{u})+\nabla \cdot (q_\parallel \mathbf{b})-2q_\perp \nabla\cdot\mathbf{b}=-2p_\parallel \mathbf{bb}\ \colon\nabla\mathbf{u}+2\nu_c\Delta p. \nonumber
\end{equation}
The equations are written in Gauss units, $\mathbf{u}$ and $\mathbf{B}$ are the ion flow velocity and magnetic field. The unit vector $\mathbf{b}=\mathbf{B}/B$ denotes the background field direction. The ion mass density is denoted as $\rho$ and the ion collision frequency $\nu_c$. The components of the pressure tensor $p_\perp$ and $p_\parallel$ \footnote{In principle, the pressure should be solved kinetically for each species but in the equivalent formalism of Kinetic-MHD only the ion pressure is considered. This approximation can be formally justified by an expansion in the electron-ion mass ratio when the electrons are moderately collisional (see Appendix A of \citet{Rosin}). In the Earth's magnetosheath, the electron pressure is small compared to the ion pressure by a factor $m_e v_{te}^2/m_i v_{ti}^2\simeq 10^{-1}$. The neglect of electron pressure in collisionless magnetosheath environments produced from large Mach number shocks might require two-fluid treatment.} are parallel and perpendicular to the magnetic field and summed over the ion and electron species. Similarly with the heat fluxes $q_\perp$ and $q_\parallel$. We define departure from isotropy with the parameter: 
\begin{equation}
\Delta(t)\equiv \frac{\Delta p}{p_0}=\frac{p_\perp-p_\parallel}{p_0}=3\frac{p_\perp-p_\parallel}{2p_\perp+p_\parallel}.
\end{equation}  
If the heat fluxes are not negligible, they need to be solved kinetically or require the use of a closure scheme \cite{hammett1992fluid, passot2012extending}. In the following, we focus on the impact of pressure anisotropies on macroscopic plasma properties and discuss the necessity to account for heat fluxes as well as pressure anisotropies in the conclusion. Our aim, is to use Equation sets (1) to determine the conditions under which kinetic energy in the plasma can be enhanced at levels comparable to or greater than the Alfv\'en speed $V_A= B_0/\sqrt{4\pi \rho}$. 

\subsection{Quantifying enhancement of kinetic energy}
The equation set (\ref{Initial_eq}) can be used to determine what mechanisms can produce enhancements in the kinetic energy density $E_k$:
\begin{equation}
E_k\equiv\int_\mathcal{V} \frac{\rho u^2}{2}dV.
\end{equation}
For closed and bounded volumes, the equation set (\ref{Initial_eq}) conserves energy, i.e., the sum of kinetic energy $E_k$, magnetic energy $E_m=\int_\mathcal{V}{B^2}/{8\pi} dV$ and thermal energy $E_{th}=\int_\mathcal{V} \left(p_\perp+{p_\parallel}/{2}\right) dV$ is constant. For a collisionless plasma ($\nu_c=0$) with negligible heat fluxes ($q_\perp=q_\parallel=0$), the kinetic energy evolves as:
\begin{widetext}
\begin{eqnarray}
\label{equation_uno}
\frac{d E_k}{d t}=&-&\int_{\mathcal{S}}\Bigg{[}\underbrace{\frac{\rho u^2}{2} \mathbf{u}}_{\mycircle{1}}+\underbrace{p_\perp\left(\mathbf{I}-\frac{p_\perp-p_\parallel}{p_\perp} \mathbf{bb}\right)\cdot\mathbf{u}}_{{\mycircle{2}}}+\underbrace{\left(\frac{B^2}{8\pi}\mathbf{I}-\frac{\mathbf{BB}}{4\pi}\right)\cdot\mathbf{u}}_{{\mycircle{3}}}\Bigg{]}\cdot d\mathbf{S}\nonumber \\ 
&+&\int_\mathcal{V}  p_\parallel\Bigg{[} \underbrace{\nabla \cdot \mathbf{u}\left(1-\frac{1}{\beta_\parallel}\right)}_{{\mycircle{4}}} +\underbrace{\frac{d\ln B}{dt}\left(1-\frac{T_\perp}{T_\parallel}-\frac{2}{\beta_\parallel}\right)}_{\mycircle{5}}\Bigg{]} d{V}.
\end{eqnarray}
\end{widetext}
The evolution of the kinetic energy is controlled by five terms annotated on the right-hand side of Equation (4). The first term annotated as $\mycircle{1}$ in the surface integral determines the rate at which kinetic energy penetrates across the surface. The second and third terms annotated as $\mycircle{2}$ and $\mycircle{3}$ determine the rate at which pressure and magnetic stresses do work on the boundary \footnote{We here envision a blob of plasma magnetically connected to the shock transition layer. The boundary in question is at the shock transition layer where kinetic energy can penetrate and where thermal and magnetic pressure can apply a tension force.}. The fourth term $\mycircle{4}$ quantifies the impact of compressive heating in terms of the parallel plasma beta $\beta_\parallel=8\pi p_\parallel/B^2$. Note how the effect of compression reverses for $\beta_\parallel=1$. The fifth term $\mycircle{5}$ describes the growth of kinetic energy density when the magnetic field amplitude $B$ grows, and when the coefficient $ 1-\frac{T_\perp}{T_\parallel}-\frac{2}{\beta_\parallel}$
is positive. These latter two conditions are met for a kinetically dominated firehose unstable plasma \cite{Chandra_Firehose, Parker_Firehose, Hunana_2017}. 

Equation (4) is instructive on multiple counts. It highlights the contribution of pressure and temperature anisotropy on kinetic energy growth, but also allows us to revisit known mechanisms. For instance, the explanation for jets due to solar wind penetration of a shock surface perturbed by coherent structures suggested by \citet{Hietala13} is contained in the first term of the surface integral 

\footnote{As noted by \citet{blandford1987particle}, a shock front can be described as a surface of discontinuity across which mass, momentum, and energy flow steadily. Although no shock wave is perfectly steady or discontinuous, this assumption holds if the variation in flow variables occurs over a distance much smaller than the corresponding scales ahead of and behind the shock. Additionally, the overall flow pattern should remain relatively unchanged during the time it takes for a fluid element to traverse the shock. While the ripple mechanism of \citet{Hietala13} can conceptually contribute to enhanced kinetic energy downstream, it relies on the assumption that the quasi-parallel shock is locally stationary on spatial and temporal scales comparable to those of SLAMS, which are specifically the structures forming the quasi-parallel transition layer. It has also been known for decades that the quasi-parallel shock produces a broad spectrum of turbulent fluctuations \cite{eastwood2005foreshock} and coherent structures, ranging from the ion gyroscale, to the curvature scale of the Earth's bow shock \cite{schwartz1992observations, wilson2013shocklets, blanco2011foreshock, schwartz1985active, schwartz2018ion, turner2013first}.  It is therefore difficult to conceive of the quasi-parallel shocks as stationary on a wide range of spatial and temporal scales. In the view of the authors, the effects due to shock temporal and spatial variability, require a perturbed study of MHD shocks. A stability theory of hydrodynamical shock waves has been developed more than half a century ago by \citet{Dyakov} and \citet{Kontorovich} for small amplitude perturbations, but is currently missing for large amplitude perturbations at MHD shocks.}. Similarly, mechanisms assuming that enhanced density and thermal pressures in the solar wind drive jets at the shock boundary \cite{Archer12, Archer13} can be assimilated to the pressure stress term in Equation (4). 

However, in the following we ignore the surface terms in Equation (4) and focus instead on quantifying the effects of compression and firehose unstable plasma in enhancing kinetic energy far from the boundaries and for a small volume $\partial \mathcal{V}$ \footnote{While we acknowledge that the boundaries confining planetary magnetosheaths can hardly be described as closed and stationary domains, we here assume that pockets of localised plasma downstream of the quasi-parallel shock can experience either compression or injection of pressure anisotropic plasma.}. For these instances, the contributions from the surface integral in Equation (\ref{equation_uno}) vanish and the evolution of the kinetic energy can be written as:
\begin{equation}
\label{Master}
\frac{d E_k}{d t} =\int_{\mathcal{V}} p_\parallel\left[ \nabla \cdot \mathbf{u}\left(1-\frac{1}{\beta_\parallel}\right) +\frac{d\ln B}{dt}\left(1-\frac{T_\perp}{T_\parallel}-\frac{2}{\beta_\parallel}\right)\right] d{V}.
\end{equation} 
Consequently, the kinetic energy can grow (decrease) for the following cases. 
\begin{itemize}
    \item \textbf{Case 1} For an incompressible plasma $\nabla \cdot \mathbf{u}={d \ln \rho}/{dt}=0$, a positive (negative) correlation between ${d\ln B}/{dt}$ and $\left(1-{T_\perp}/{T_\parallel}-{2}/{\beta_\parallel}\right)$ can result in transfer of thermal to (from) kinetic energy. This coefficient is positive during the growth of firehose instabilities, that is when $\left(1-{T_\perp}/{T_\parallel}-{2}/{\beta_\parallel}\right)>0$ and ${d\ln B}/{dt}>0$. When the plasma reaches firehose marginal stability and/or when the magnetic field amplitude saturates, the kinetic energy growth ceases. 
    \item \textbf{Case 2} For a firehose stable, but compressible plasma, kinetic energy density can grow for positive (negative) correlations between, $\nabla \cdot \mathbf{u}$ and  $\left(1-\frac{1}{\beta_\parallel}\right)$. Hence for low plasma $\beta_\parallel (<1)$, compression ($\nabla \cdot \mathbf{ u} <0$) results in kinetic energy density enhancement, whereas for high $\beta_\parallel( >)1$, rarefaction ($\nabla \cdot \mathbf{u}>0$) can lead to such an enhancement.
\end{itemize}
In the following we prescribe background flow properties to determine the conditions under which the kinetic energy can grow to levels consistent with jets' observations.

\subsection{Enhancement in kinetic energy density due to adiabatic compression and rarefaction} 
We are interested with the case of a pure compressible background flow $\mathbf{u}_0$ devoid of shear and its associated impact upon the kinetic energy for a firehose stable plasma. We write the mean flow as ${u}_{0i}=A_{ij} {x_j}=\lambda(t)\delta_{ij}{x_j}$, with the diagonal matrix $A_{ij}$ written in terms of the Kronecker delta $\delta_{ij}$ and a time-varying compression rate $\lambda(t)=\dot{L}/L$. If the compression is anisotropic we can define $\lambda_\perp(t)=\dot{L}_\perp/L_\perp$ and $\lambda_\parallel(t)=\dot{L}_\parallel/L_\parallel$ and write the matrix $A_{ij}$ as:
\[A_{ij}=
\begin{pmatrix}
  \lambda_\perp & 0 & 0 \\
  0 & \lambda_\perp & 0 \\
  0 & 0 & \lambda_\parallel
 \end{pmatrix}
 \]
 or equivalently $\mathbf{A}=\lambda_\perp(\mathbf{I-bb})+\lambda_\parallel\mathbf{bb}$. Using the set of Equations (\ref{Initial_eq}) with zero heat fluxes in the collisionless limit, it is easy show that the background flow quantities evolve as follow:
 \begin{equation}
 \rho(t)=\rho_0 \left(\frac{L_{0\perp}}{L_\perp}\right)^2\left(\frac{L_{0\parallel}}{L_\parallel}\right)
 \end{equation}
 \begin{equation}
 B_0(t)=B_0(0) \left(\frac{L_{0\perp}}{L_\perp}\right)^2
 \end{equation}
  \begin{equation}
 p_\parallel(t)=p_\parallel(0) \left(\frac{L_{0\perp}}{L_\perp}\right)^4\left(\frac{L_{0\parallel}}{L_\parallel}\right)^3
 \end{equation}
  \begin{equation}
 p_\perp(t)=p_\perp(0) \left(\frac{L_{0\perp}}{L_\perp}\right)^4\left(\frac{L_{0\parallel}}{L_\parallel}\right)
 \end{equation}
  \begin{equation}
    \label{betapar}
 \beta_\parallel(t)=\beta_\parallel(0) \left(\frac{L_{0\perp}}{L_\perp}\right)^4\left(\frac{L_{0\parallel}}{L_\parallel}\right)^3
 \end{equation}
  \begin{equation}
  \label{Tpar}
 T_\parallel(t)=T_\parallel(0) \left(\frac{L_{0\perp}}{L_\perp}\right)^2\left(\frac{L_{0\parallel}}{L_\parallel}\right)^2
 \end{equation}
  \begin{equation}
    \label{Tperp}
 T_\perp(t)=T_\perp(0) \left(\frac{L_{0\perp}}{L_\perp}\right)^2
 \end{equation}
 \begin{equation}
 V_A(t)=V_A(0) \left(\frac{L_{0\perp}}{L_\perp}\right)\left(\frac{L_{\parallel}}{L_{0\parallel}}\right)^{1/2}.
 \end{equation}
We note that the above solutions are consistent with the Chew-Goldberd-Low invariants $p_\perp/\rho B$ and $p_\parallel  B^2/\rho^3$  \citep{CGL} for compression ($L_{\perp,\parallel}(t)\leq L_{\perp,\parallel}(0)$) and rarefaction ($L_{\perp,\parallel}(t)\geq L_{\perp,\parallel}(0)$). We now assume an incompressible perturbation $\nabla\cdot \mathbf{u}_{\perp1}=0$ and compute the rate of change of kinetic energy density for the low Mach number regime \footnote{In the low Mach number regime we can neglect the density fluctuations, i.e. $\rho_1\ll \rho_0$}. Under these assumptions the rate of change of kinetic energy density can be written as:
\begin{eqnarray}
\label{compressible_eq}
\frac{\partial}{ \partial t} \left(\frac{\rho u^2}{2}\right)&=&\frac{B^2}{8\pi} \nabla \cdot \mathbf{u}_0\left(\beta_\parallel-1\right)\nonumber\\ 
&=&\frac{B(0)^2}{8\pi}\int dV  \left(2\frac{\dot{L}_\perp}{L_\perp}+\frac{\dot{L}_\parallel}{L_\parallel}\right)\left(\frac{L_{0\perp}}{L_\perp}\right)^4 \nonumber \\
&\times&
\left[\beta_\parallel(0))\left(\frac{L_{0\perp}}{L_\perp}\right)^4\left(\frac{L_{0\parallel}}{L_\parallel}\right)^3- 1\right]
\end{eqnarray}
For the sake of simplicity, we also set $\dot{L}_\parallel=0$ and assume a compression evolving linearly in time, i.e., $L_\perp(t)=L_\perp(0)-U_b t$ with $U_b>0$. Equation (\ref{compressible_eq}) can then be written as:
\begin{eqnarray}
\frac{\partial}{ \partial t} \left(\frac{\rho u^2}{2}\right)=-\frac{B(0)^2}{4\pi}\frac{U_b}{L_0}\left(\frac{L_0}{L_\perp}\right)^5\left[\beta_\parallel(0)\left(\frac{L_{0}}{L_\perp}\right)^4-1\right]. \nonumber
\end{eqnarray}
The kinetic energy density normalized by the magnetic energy density, i.e., $\widetilde{E_k}=8\pi E_k/B(0)^2$, during compression of the plasma, can be obtained from integration of the above expression:
\begin{eqnarray}
\widetilde{E_k}(t)-\widetilde{E_k}(0)&=&-2\frac{U_b}{L_0}\int_{0}^t dt'\left(\frac{L_0}{L_\perp(t')}\right)^5 \nonumber \\
&\times&\left[\beta_\parallel(0)\left(\frac{L_{0}}{L_\perp(t')}\right)^4-1\right]
\end{eqnarray}
We now make the following change of variables $\tau \longrightarrow |U_b|t'/L_0$ and $L(t')\longrightarrow L(t')/L_0=1-\tau$. The resulting integral can then be written as:
\begin{eqnarray}
\widetilde{E_k}(t)-\widetilde{E_k}(0)&=&-2\int_{0}^{\tau} d\tau'\left(\frac{1}{1-\tau'}\right)^5\nonumber \\
&\times&\left[\beta_\parallel(0)\left(\frac{1}{1-\tau'}\right)^4-1\right] \\
&=&\frac{\beta_\parallel(0)}{4}\left[1-\frac{1}{(1-\tau)^8}\right]+\frac{1}{2}\left[\frac{1}{(1-\tau)^4}-1\right]. \nonumber
\end{eqnarray}
We can therefore compute the resulting enhancement in kinetic energy density as a function of the initial value of $\beta_\parallel$ and time. In Figure 1, we show the kinetic energy $\widetilde{E_k}$ evolution as a function of time $\tau$ for density compressions of $\rho(t)/\rho(0)\leq 2$. The lower the initial plasma beta the higher the enhancement in kinetic energy density. We note that this enhancement in kinetic energy density is physically consistent with the study of \citet{Lavraud} for kinetic energy density reaching values of the order of the magnetic energy for low plasma beta. For plasma conditions typical of the Earth's magnetosheath in proximity to the magnetopause boundary, kinetic energy density initially in partition with magnetic energy density, i.e., $\widetilde{E_k}(0)=1$, can be amplified into super-Alfv\'enic flows $u$ of the order of $ 5 B_0/\sqrt{8\pi \rho}\simeq 3.5 V_A(0)\simeq 2 V_A\sim 300-400$ km/s. However, this adiabatic compression follows the CGL relations and results in a growth of the plasma beta towards $\beta_\parallel \simeq 1$. When the plasma beta reaches a value of 1 due to compression the growth stops and reverses. Such a compressive mechanism can therefore explain transient enhancement of kinetic energy density in low plasma beta. 

As noted in the previous section, rarefaction ($\nabla\cdot\mathbf{u}>0$) for a high plasma beta $\beta_\parallel>1$ can also result in enhanced kinetic energy density. Assuming once more an adiabatic flow that expands linearly, as per $L_\perp(t)=L_\perp(0)+U_b t$ with $U_b>0$ and $\dot{L}_\parallel=0$, we can compute the rate of change of kinetic energy density. Using Equation (\ref{compressible_eq}) we find:
\begin{eqnarray}
\widetilde{E_k}(t)-\widetilde{E_k}(0)
&=&\frac{\beta_\parallel(0)}{4}\left[1-\frac{1}{(1+\tau)^8}\right]+\frac{1}{2}\left[\frac{1}{(1+\tau)^4}-1\right]. \nonumber
\end{eqnarray}
with the solution plotted for various values of initial plasma beta in Figure 2. The kinetic energy density grows monotonically, but as the plasma beta reduces to $\beta_\parallel \simeq 1$, the kinetic energy density saturates. As $\tau \gg 1$, the normalised kinetic energy density saturates at a value of $\widetilde{E_k} \simeq \frac{\beta_\parallel}{4}-\frac{1}{2}$. Thus, for even moderate plasma beta values of 10 which are commonly found in the magnetosheath \cite{dimmock2015statistical}, the kinetic energy density can grow by a factor of 2 to 3 with respect to the magnetic field energy density and result in super-Alfv\'enic flows. We therefore close this secton by pointing out that compressed and rarefied plasma can both be associated with adiabatic enhancement in the kinetic energy density. And as a result of compression for low plasma beta and rarefaction in high beta plasma, the resulting plasma beta parameter reaches values of $\beta_\parallel \simeq 1$, which is consistent with observation of jets \cite{raptis2020classifying}.

\begin{figure}
\includegraphics[height=8.cm,width=8.3cm, keepaspectratio]{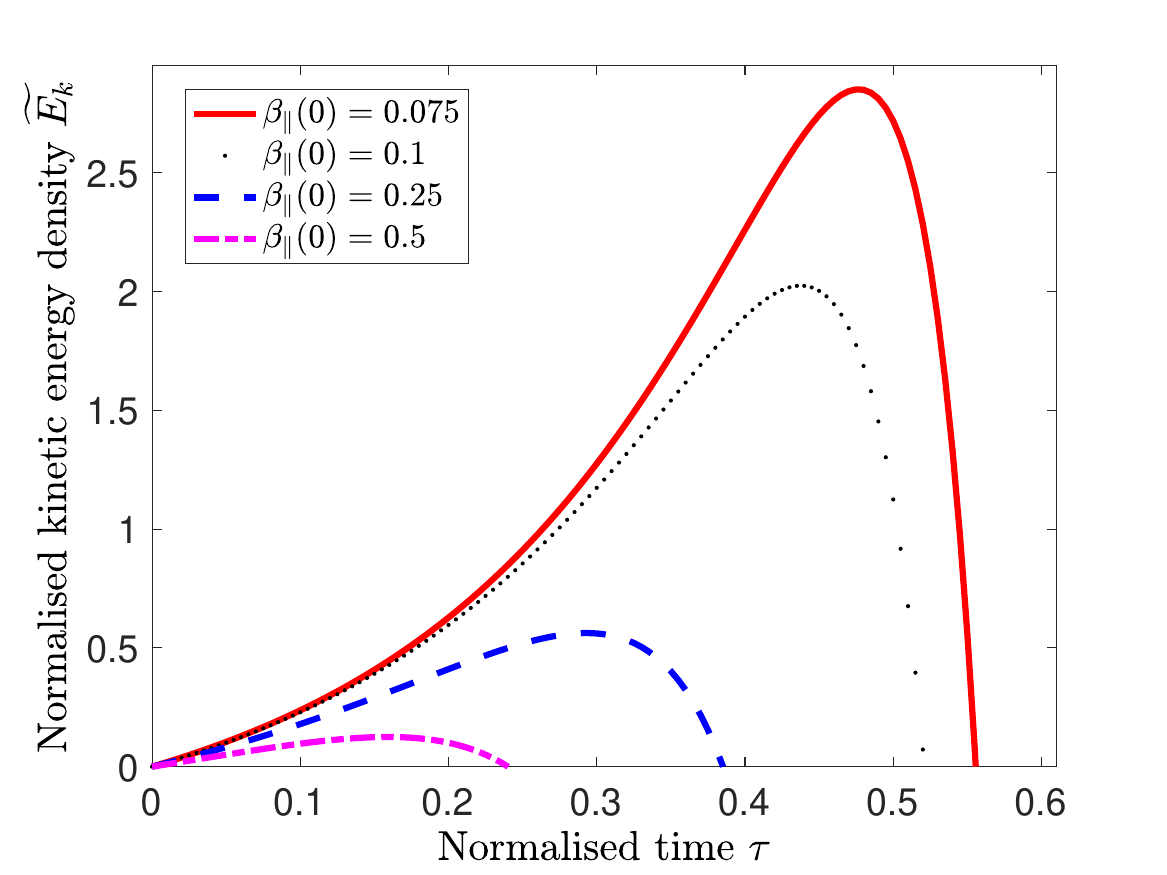}
\caption{Kinetic energy density normalized by the magnetic field energy density as a function of normalised time $\tau=|U_b| t/L_0$ for a compressible flow. Curves are plotted for $\beta_\parallel(0)$ varying between 0.02 and 0.1. Larger kinetic energy densities are reached for lower initial $\beta_\parallel$ value. The case study presented by \cite{Lavraud} corresponds to such instance of kinetic energy density enhancement for low $\beta_\parallel \ll 1$. This figure indicates that the resulting flow can become super-Alfv\'enic with values of the order of 2-3 times the Alfv\'en speed.}
 \end{figure} 

 \begin{figure}
\label{figcomp1}
\includegraphics[height=8.cm,width=8.3cm, keepaspectratio]{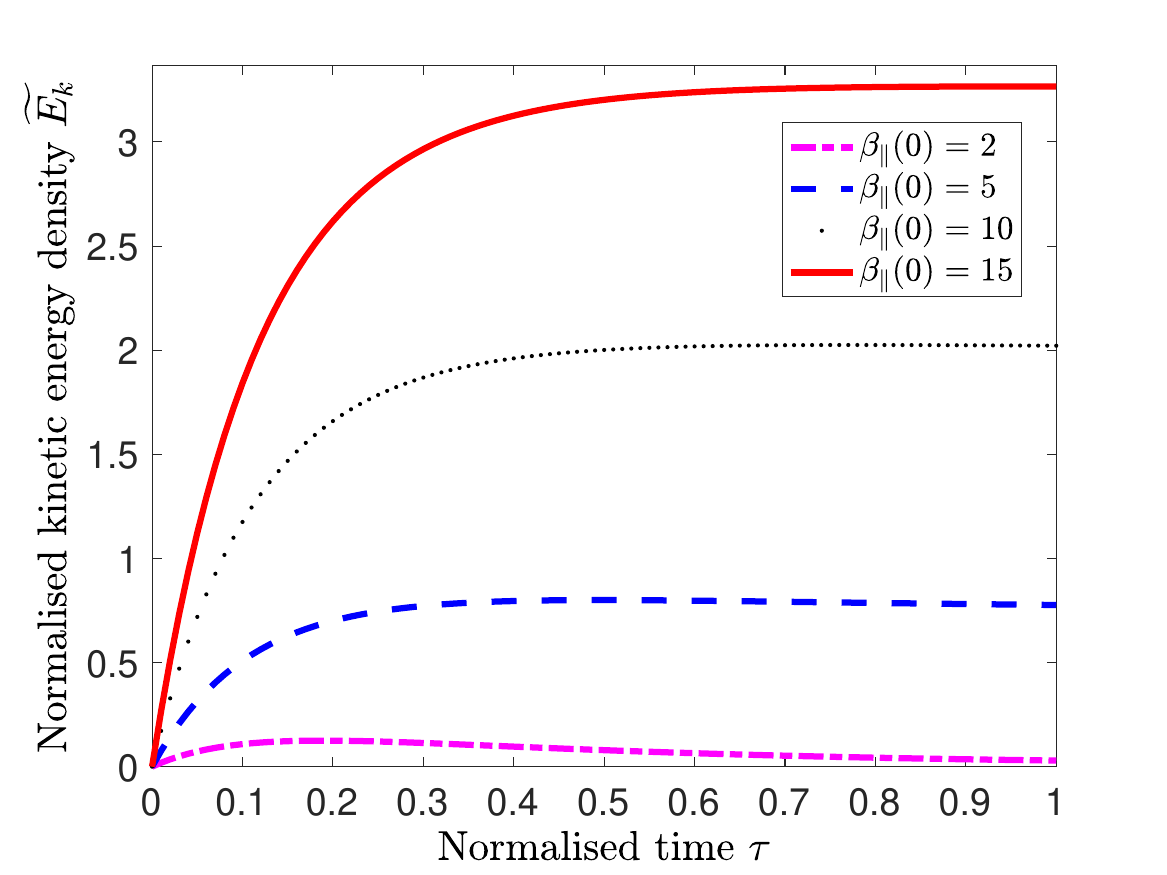}
\caption{Kinetic energy density normalized by the magnetic field energy density as a function of normalised time $\tau=|U_b| t/L_0$ for an expanding flow. Curves are plotted for $\beta_\parallel(0)$ varying between 2 and 12. Larger kinetic energy densities are reached for higher $\beta_\parallel$ value. This figure indicates that high plasma beta rarefied flows can become super-Alfv\'enic with values of the order of 2-3 times the Alfv\'en speed. }
 \end{figure}

\subsection{Enhancement in kinetic energy due to firehose unstable plasma}
Plasma environments downstream of astrophysical shocks are fertile grounds for the growth of pressure-driven instabilities. In the effectively collisionless magnetosheath of the Earth one can think of at least three different mechanisms to sustain pressure anisotropies: (1) locally at the shock transition due to compression and anisotropic ion heating; (2) in the inner magnetosheath by (mesoscale) turbulent fluctuations and (3) closer to magnetopause boundary by large-scale field inhomogeneities. In the following we assume that the pressure anisotropy is driven at the shock boundary, since jets are observed immediately downstream of the Earth's bow shock. To estimate the scale of the pressure anisotropy downstream of collisionless plasma shock we assume the applicability of the Rankine-Hugoniot equation. With the Rankine-Hugoniot relations, one can estimate the associated temperature or pressure anisotropy as a function of Mach number for parameters consistent with the Earth's quasi-parallel shock. The estimated values for the firehose instability criteria downstream of the Earth's bow shock are shown in Figures 5 of the Appendix. We here infer from the Rankine-Hugoniot conditions that quasi-parallel shocks ($\theta_{Bn}<30^o$ and $4<M_A<20$) can seed downstream plasmas with temperature anisotropies $T_\perp/T_\parallel \simeq 0.1-0.5$ and firehose instability criteria $|\Delta_0 +2/\beta_\parallel| \simeq 0.1-1$\footnote{We note that higher values in the firehose instability criteria are theoretically possible according to Rankine-Hugoniot relations, but effectively impossible to be observed \textit{in situ} since the associated instability growth rate is too fast \cite{kennel1967collisionless, schekochihin2010magnetofluid}. Observationally, what is measured is the saturated anisotropy level \cite{bale2009magnetic, dimmock2015statistical}. In terms of modelling, such large instability criteria can not be treated through perturbation methods, as done in \cite{Rosin}, and require more advanced theoretical and numerical studies.}. 

In the instance where the plasma experiences adiabatic compression, we can use equations (\ref{betapar}), (\ref{Tpar}) and (\ref{Tperp}) to determine when the stability criteria for the parallel firehose ($k_\perp=0$) is violated:
\begin{eqnarray}
\label{instability_criterion}
1-\frac{T_\perp(t)}{T_\parallel(t)}-\frac{2}{\beta_\parallel(t)}&=&1-\frac{T_\perp(0)}{T_\parallel(0)}\left(\frac{L_{\parallel}}{L_{0\parallel}}\right)^2\nonumber \\
&-&\frac{2}{\beta_\parallel(0)}\left(\frac{L_{\parallel}}{L_{0\parallel}}\right)^3\left(\frac{L_{\perp}}{L_{0\perp}}\right)^4>0. \nonumber
\end{eqnarray}
It is clear from the above equation that even if the plasma is initially stable to firehose instability, compression drives the plasma to the firehose unstable threshold. Similarly, an initially firehose unstable plasma can be made stable by rarefaction, and in the process trigger the mirror instability. Using the parameter regime of the previous section ($\dot{L}_\parallel=0$, $\beta_\parallel(0) \simeq 0.1-1$) and $T_\perp(0)/T_\parallel(0)=0.75-0.95$, Equation (\ref{instability_criterion}) provides us with a density compression of order 3/2 to 4 and a time $\tau \simeq 1/2$ for the firehose instability to be triggered. Thus, firehose instability can be sustained through locally induced compressive fluctuations commonly found in the turbulent magnetosheath \cite{sahraoui2020magnetohydrodynamic}.

In order to quantify the amplification of the perturbed flow $\delta \mathbf{u_\perp}$ arising from the triggering of the firehose instability, either through adiabatic compression or non-adiabatic heating at the shock, we use the asymptotic theory of \citet{Rosin} derived for the case where the instability could be driven by shear and compression of the mean flow. In the absence of collisions, the anisotropy for the parallel firehose evolves according to:
\begin{equation}
\Delta(t)=\Delta(0)+3\int_0^t dt' \gamma(t')+\frac{\overline{\delta B_\perp^2}}{B_0^2}, \nonumber
\end{equation}
where $\gamma$ is the drive rate \footnote{The drive rate in astrophysical plasmas, whether originating in compressive or shear motion, is related to the change in magnetic field amplitude, i.e., $\gamma\sim d \ln B/dt=\mathbf{bb}: \nabla \mathbf{u-\nabla\cdot u}$. } bringing the anisotropy to unstable levels, and the bar above the perturbed normalized magnetic energy denotes an average over fast time-scales. Saturation occurs when $\Delta(t)\simeq -2/\beta_\parallel$ after magnetic field fluctuations locally reduce the plasma beta. Using the asymptotic construction of \citet{Rosin}, we integrate the firehose equations for the perturbed flow velocity:
\begin{eqnarray}
\frac{\partial}{\partial t}\delta \mathbf{u_\perp}=\frac{v_{thi}^2}{2}\nabla_\parallel\left[\left(\Delta(t)+\frac{2}{\beta}\right)\frac{\delta \mathbf{B}_\perp}{B_0}+\frac{\nabla_\parallel \delta\mathbf{u}_\perp}{\Omega_i}\times \mathbf{b}\right] \nonumber
\end{eqnarray}
and the perturbed magnetic field:
\begin{equation}
\frac{\partial}{\partial t} \left(\frac{\delta \mathbf{B_\perp}}{B_0}\right)=\nabla_\parallel \delta \mathbf{u_\perp}. \nonumber
\end{equation}
In the above $v_{thi}$ stands as the ion thermal velocity, and $\Omega_i$ as the ion gyrofrequency. We set $\beta_i=10$, $\nu/\Omega_i=10^{-3}$ and integrate for a single mode of scale $k\rho_i=0.1$ for a time $\Omega_it=10^3$, i.e., before one collision time. Figure (3) shows the amplitude of the normalized perturbed flow for three initial pressure anisotropies $\Delta(0)=[-0.1, -0.3, -0.5]$ consistent with Rankine-Huogoniot estimates. We note that for $\Delta(0)=-0.1$ the flow perturbation is modest with ${\delta u_\perp}/u_0\leq 10^{-3}$. On the other hand, for $|\Delta(0)|\geq 0.3$, appreciable flow perturbation of the order of $\delta u_\perp/u_0\simeq 0.1-1$ are found. Such flow enhancements associated with large temperature anisotropies would therefore appear as kinetic scale jets in the magnetosheath. However, it should be kept in mind that when perturbation becomes comparable to the background, the asymptotic theory of \citet{Rosin} formally breaks down and additional theoretical and numerical work is needed to probe the contribution of shocked plasmas far from marginal stability. Nonetheless, our results provide evidence that firehose fluctuations can produce enhancement in kinetic energy density downstream of shocks.  

\begin{figure}
\label{fig:firehose}
\includegraphics[height=8.cm,width=8.3cm, keepaspectratio]{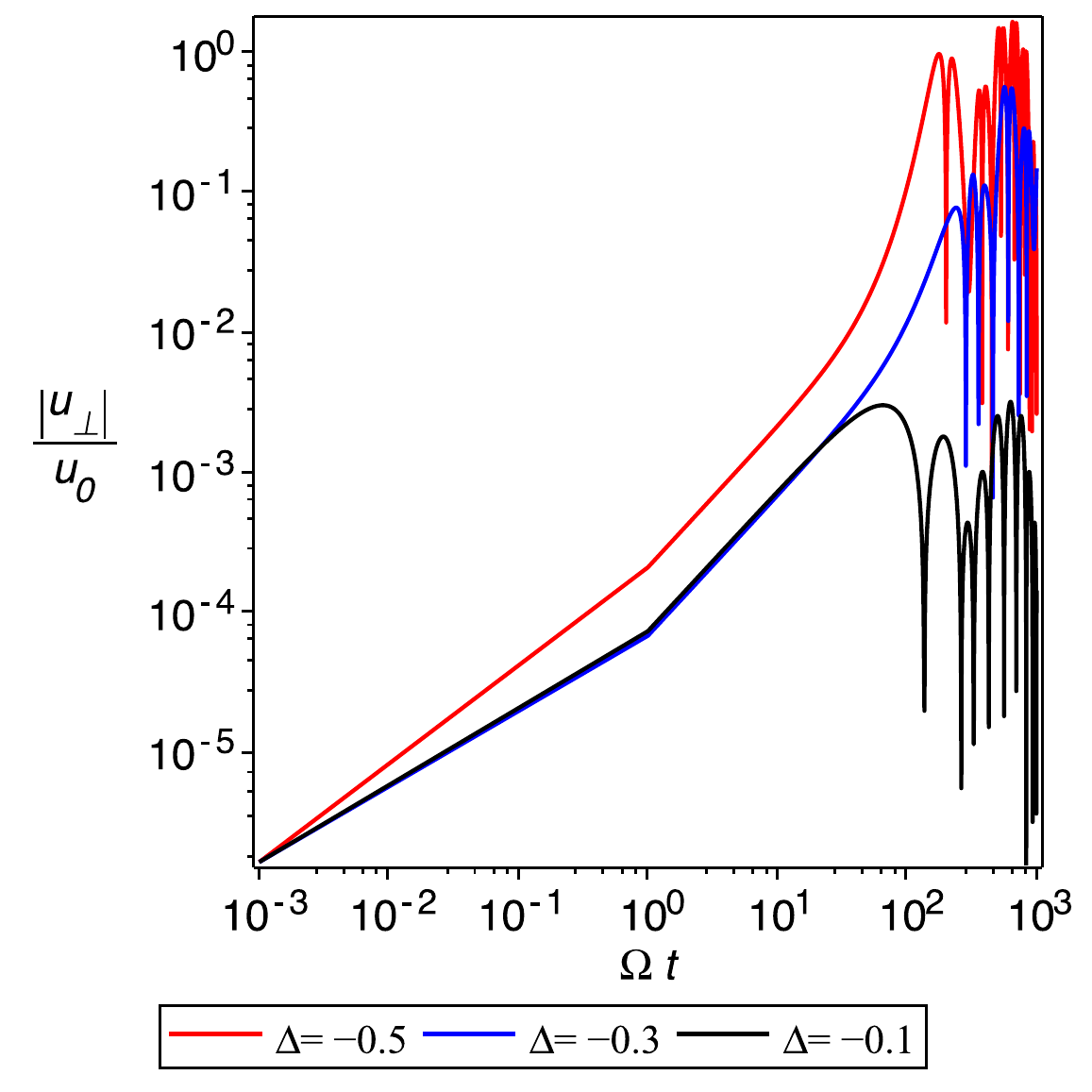}
\caption{Evolution of the perturbed flow associated with a firehose instability for three initial pressure anisotropies $\Delta(0)=[-0.1, -0.3, -0.5]$. The time is normalised by the Larmor frequency. It takes $t\simeq 100/\Omega \simeq 16$ Larmor periods $T=2\pi/\Omega$ for the flow to become comparable to the background flow.}
 \end{figure}

\section{Initial Observational Support}
\subsection{Dataset}

To evaluate the theoretical approach described above regarding the firehose unstable plasma and provide an initial observational validation, we used measurements from the Magnetospheric Multiscale (MMS) mission \cite{burch2016magnetospheric}. MMS is particularly well-suited for this type of analysis, as it can provide sub-second resolution particle moments and distributions in burst mode, allowing us to obtain measurements that are comparable to the maximum growth rates of the instabilities \cite{kennel1967collisionless, schekochihin2010magnetofluid}. We employed the Fluxgate Magnetometer (FGM) instrument with a time resolution of 0.0625 seconds \cite{russell2016magnetospheric}. For ion particle moments, we used the Fast Plasma Investigation (FPI) instrument, which provides particle moments in burst time resolution of 0.15 seconds \cite{pollock2016fast}.

\subsection{Jet Observations \& Calculations}

To evaluate jets downstream of the Earth's bow shock, we used an expanded dataset from \citet{raptis2020classifying}. Specifically, we selected a subset of a magnetosheath jet list formed out of five years of \textit{in situ} MMS observations (05/2015 - 06/2015). In this dataset, jets are defined using the typical criterion of \textit{in situ} studies, where a jet must have a dynamic pressure ($P_{dyn} = \rho u_i^2$), where $u_i$ is the ion velocity, exceeding twice the background magnetosheath value \cite{Archer12,Archer13,plaschke2018jets,raptis2020classifying}. This criterion can be expressed as:

\begin{equation}\label{jet_criterion}
    P_{jet} \geq 2 \langle P_{msh} \rangle_{20 \min},
\end{equation}
where $\langle P_{msh} \rangle_{20 \min}$ represents a moving average window of 20 minutes across the magnetosheath time series observations. Jets are initially found using the low-resolution measurements of MMS ("fast" mode - 4.5 seconds) and then characterised as a separate subset which contains burst mode data. More information regarding the dataset and its open access availability can be found in the associated open-access dataset \cite{savvas_raptis_2022_7085778}.

For our analysis, we selected a suitable set consisting of 85 quasi-parallel magnetosheath jets observed near the Earth's bow shock, for which burst data were available throughout the entire duration of the jet observation (i.e., all points that satisfy Equation (\ref{jet_criterion})), resulting in about 16300 data points. Using ion moments from the FPI instrument and resampled magnetic field data to match the resolution of the particle instrument, we calculated the temperature anisotropy ($T_\perp/T_\parallel$) and $\beta_\parallel$.

Figure 4 presents all jet-associated observations in a scatter plot, along with the thresholds for oblique firehose and mirror instability \cite{bale2009magnetic}. As demonstrated, approximately $\sim11\%$ of data points within jet observations are considered firehose unstable, with the majority of jets (60\%) exhibiting at least two data points that are unstable. Expanding these statistics to jets throughout the whole quasi-parallel magnetosheath yields similar results. It should be noted that this observational test ought to be treated as a lower threshold and as preliminary evidence that the presented theoretical framework in terms of firehose-unstable generation is applicable to multiple cases of magnetosheath jets. A more in-depth analysis is required for the following reasons. (1) Jets are expected to reach marginal stability as they get convected away from the shock transition layer and propagate into the magnetosheath. This dependence on the magnetosheath location can be quantified in a more detailed observational study. And (2), when the firehose instability is triggered far from marginal stability levels, which takes place when $T_\perp/T_\parallel \ll 1$ and $\beta_\parallel \gg 1$, the resulting fluctuations can push the plasma well above the stability threshold \cite{bott2021adaptive}. Thus, some of the fluctuations in Figure 4 that are above the marginal stability level could also have originated from firehose unstable sources. However, in order to quantify this effect, one needs to run numerical simulations of pressure-driven instabilities far from marginal stability that can account for the oblique firehose ($k_\perp \neq 0$) at inertial and Larmor scales and the fluid firehose at macroscopic scales. Future observational efforts should focus on case and statistical studies of measurements taken exceptionally close to the shock transition layer, with an evaluation of the stability criteria that accounts for large departure from marginal stability.

 \begin{figure}\label{fig:MMS}
\includegraphics[height=8.cm,width=8.3cm, keepaspectratio]{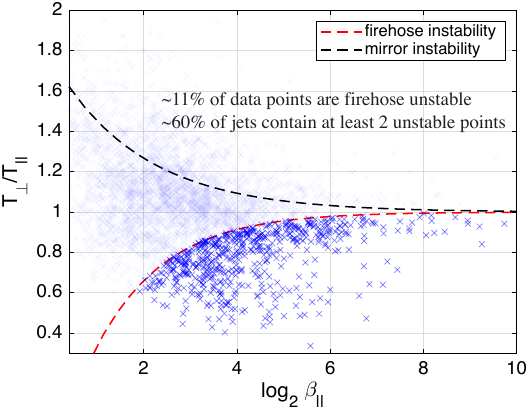}
\caption{MMS observations of quasi-parallel magnetosheath jets close to the bow shock using high resolution burst measurement as shown in temperature anisotropy ($T_\perp/T_\parallel$)  
 versus $log_2(\beta_\parallel)$ scatter plot. Instability thresholds for oblique firehose instability (lower red dotted line) and the mirror instability threshold (top dotted black line) are shown \cite{bale2009magnetic}. Each mark represents one data point of 0.15s resolution. The non-transparent points are below the firehose instability threshold ($\sim 11\%$).}
 \end{figure} 
\section{Conclusion}

Magnetosheath jets, characterized by enhanced kinetic energy density, are frequently generated downstream of Earth’s bow shock \cite{nvemevcek1998transient, Savin08, Hietala09}. Decades after the first observation of jets, the underlying mechanisms driving their formation and the role of kinetic processes remain an unresolved issue. In this report, we have used a macroscopic plasma model that incorporates kinetic processes to determine the plasma conditions under which kinetic energy density can be amplified. 

We have shown that for $\beta<1$, plasma compression results in enhancement in kinetic energy density. This instance corresponds to the case previously presented by \citet{Lavraud} in which abnormally low beta plasma ($\beta\simeq 0.1$) downstream of the shock results from the interaction of a magnetic cloud with the Earth's magnetosheath. As the plasma is being compressed near the boundary, flows of the order of a few Alfv\'en speed $u\simeq 2{V_A}\simeq 300-400$ km/s are generated. While these values are below the flow of the order of $900$ km/s observed by \citet{Lavraud}, it is computed for modest adiabatic compression of the order of $\delta \rho/\rho\leq  2$, and do not take into account boundary effects found in Equation (\ref{equation_uno}). 

In high plasmas beta, two mechanisms can lead to enhanced kinetic energy density. In the case of plasma expansion, flows become super-Alfv\'enic under typical magneotsheath plasma beta conditions of $\beta_\parallel\simeq 10$ \cite{dimmock2015statistical}. For firehose-unstable plasmas with $\beta>1$, which are more likely to occur downstream of the quasi-parallel shock \cite{yordanova2020current}, it was shown, using the asymptotic model of \citet{Rosin} that kinetic energy can grow to levels comparable to the background flow, and therefore results in doubling the flow speed. Similarly, as for the case with $\beta<1$, the resulting flows triggered by firehose instabilities are below the largest observed values. However, larger pressure anisotropies should result in larger energy deposition in the kinetic energy density and in magnetic fluctuations, but such theoretical and numerical study currently lies beyond our current understanding of firehose instability \footnote{The derivation of \citet{Rosin} also neglects the compressibility of the firehose fluctuations. This neglect is justified for $\beta \gg 1$ and $\delta B_\perp/B_0 \ll 1$. These two conditions are seldom respected in the magnetosheath, a region formed as the result the compression of the solar wind, and where compressible fluctuations can scale as $\delta \rho \sim \rho_0$.}. 

Similar to the ripple mechanism proposed by \citet{Hietala13}, the high plasma beta mechanism associated with firehose instability discussed in this report aligns with the conditions that favor jet formation, specifically that high-speed jets predominantly occur downstream of the quasi-parallel shock when
the interplanetery magnetic field and the normal vector of the shock have small angles ($\theta_{Bn}<45$). However, the firehose mechanism offers an additional advantage by explaining multiple jet properties that taken in combination are inconsistent with the ripple scenario. For example, \citet{raptis2020classifying} found that quasi-parallel jets have lower plasma beta compared to the surrounding magnetosheath plasma, while \citet{Plaschke13} demonstrated (see Figure 9) that jets are more isotropic than the surrounding magnetosheath plasma.  A plasma that is warmer than the solar wind but exhibits a lower beta parameter and reduced anisotropy compared to the surrounding magnetosheath is consistent with the saturation of firehose fluctuations downstream of the quasi-parallel shock. 

Finally, another kinetic source that can sustain enhanced kinetic energy density—or equivalently, trigger pressure anisotropies leading to jet formation—is heat fluxes. Observational studies have shown for decades that strong field-aligned heat flux persists across the entire magnetosheath, from the magnetopause to the bow shock \cite{liu2024magnetosheath}, making the production of heat flux instabilities in the high plasma beta magnetosheath plausible \cite{Rosin, schekochihin2010magnetofluid}. With recent observational evidence indicating that jets occur on smaller scales than previously reported \cite{plaschke2020scale, raptis2022magnetosheath}, our results provide new justification for the inclusion of kinetic processes in understanding the generation and evolution of magnetosheath jets. Future research should explore the role of heat fluxes, boundary stresses, and the specific conditions under which plasma compression and instabilities combine to produce super-Alfv\'enic jets.

\begin{acknowledgments}
AO expresses his gratitude to Heli Hietala for helpful  discussions and insights which contributed to the development of this work. Support for AO was provided by the Research Council of Finland profiling action Matter and Materials (grant \# 318913). SR is supported by the Magnetospheric Multiscale (MMS) mission of NASA’s Science Directorate Heliophysics Division via subcontract to the Southwest Research Institute (NNG04EB99C). SR acknowledges the support of the International Space Sciences Institute (ISSI) team 555, “Impact of Upstream Mesoscale Transients on the Near-Earth Environment” SR also acknowledges funding from National Science Foundation (NSF) Geospace Environment Modeling (GEM) program 2225463.
\end{acknowledgments}

\appendix

\section{Rankine-Hugoniot Estimate of Temperature Anisotropy}
In order to estimate the shock conditions required to sustain firehose unstable plasmas, we solve the Rankine-Hugoniot equations that incorporate temperature anisotropies for the shock-angle $\theta_{Bn}$, the angle between the interplanetary magnetic field and the shock normal. The Rankine-Hugoniot relations in the shock frame are given by \citet{liu2007temperature} and since the equations are undetermined, we need to fix one parameter to solve them. We follow the approach of \citet{liu2007temperature} and fix the ratio of downstream to upstream density $r=\rho_2/\rho_1$. The numerical solutions are shown in Figure 5 for the temperature anisotropy and Figure 6 for the firehose instability criteria. We note that according to the Rankine-Hugoniot relations, the firehose instability criterion is larger for high Mach number quasi-parallel shocks, but that when the Mach number is less than 6, firehose unstable plasma can be generated for both quasi-parallel and quasi-perpendicular shocks.  

\begin{figure*}
\includegraphics[height=8.3cm,width=10.cm, keepaspectratio]{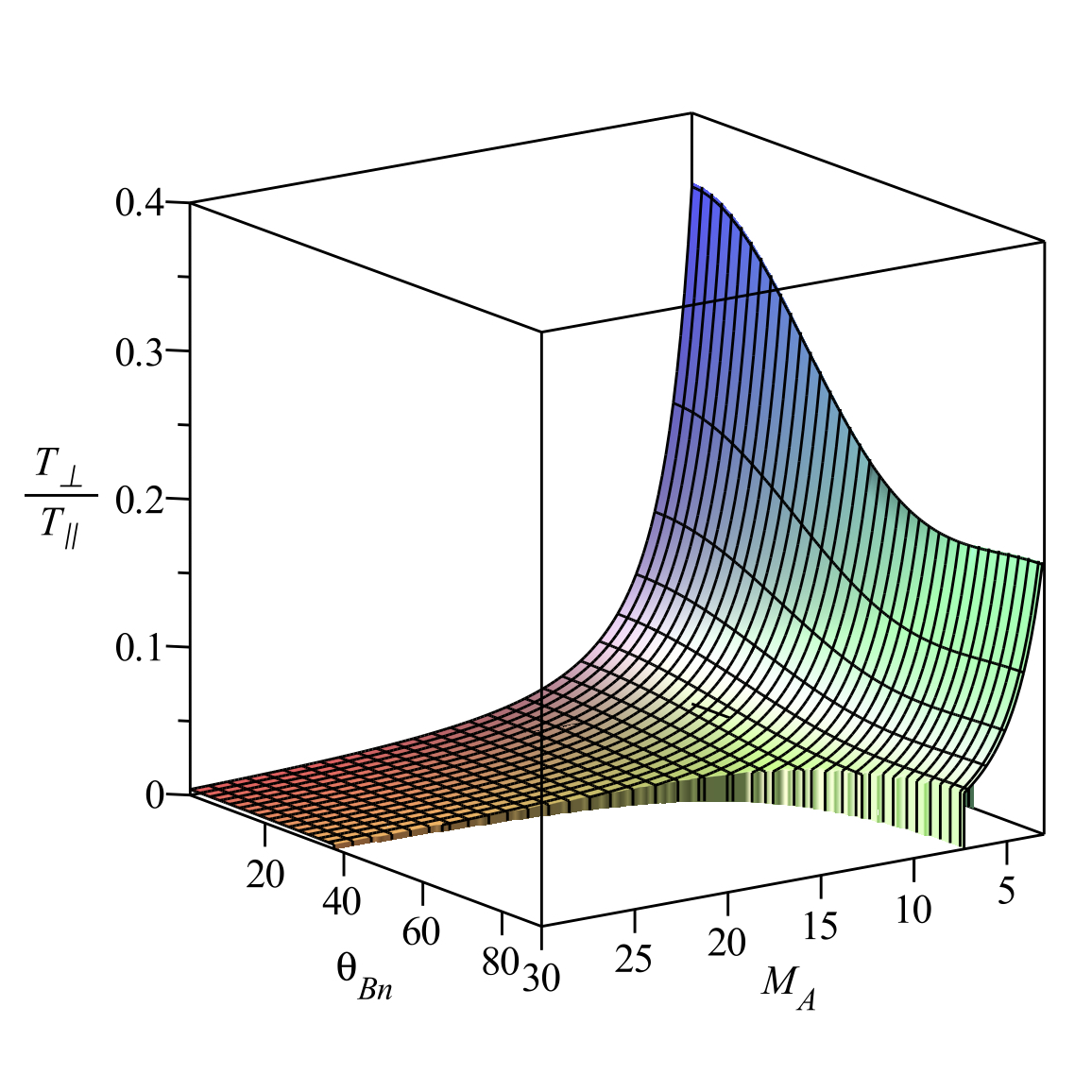}
\includegraphics[height=8.3cm,width=10.cm, keepaspectratio]{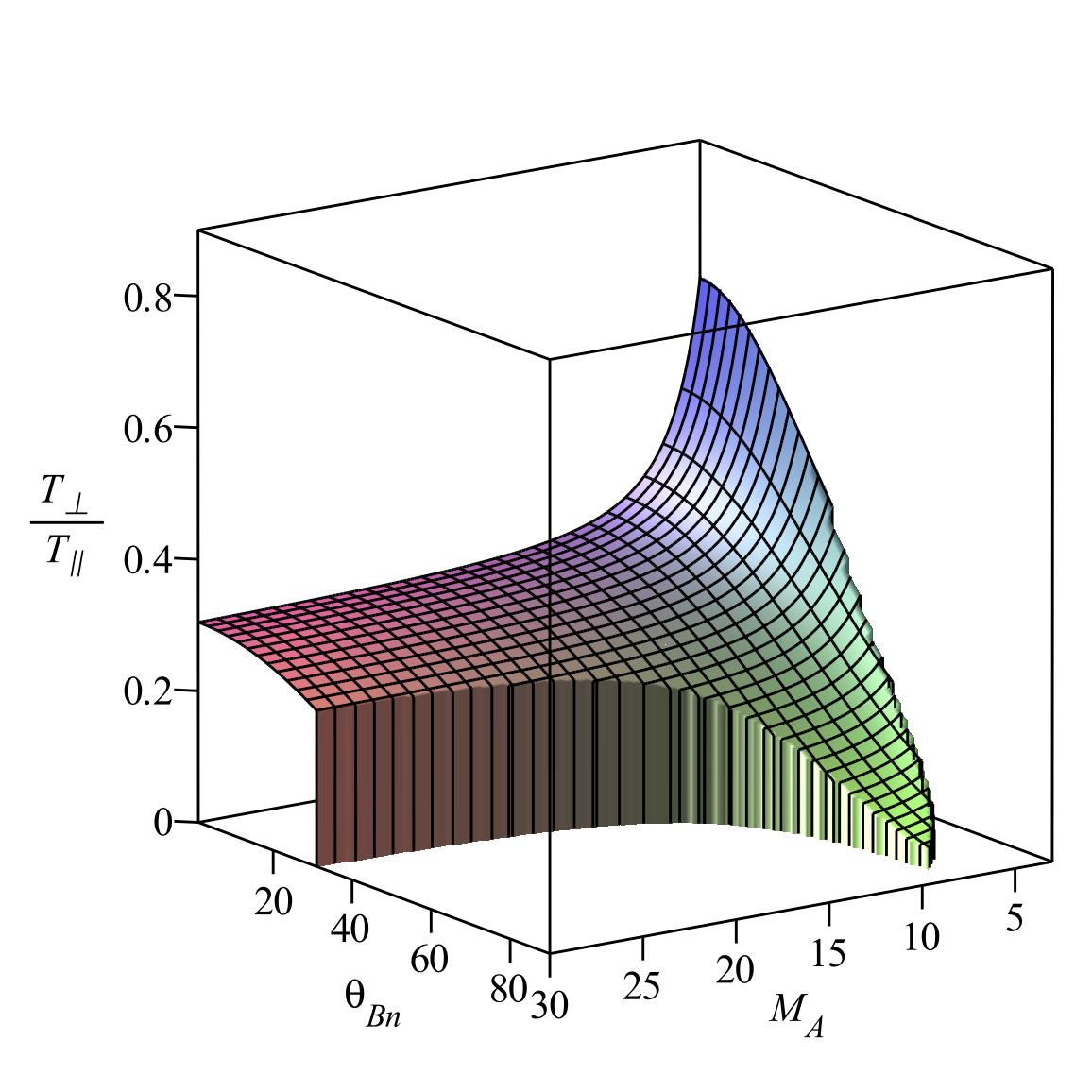}
  \caption{Surfaces for the temperature anisotropy $T_\perp/T_\parallel>0$ as a function of $\theta_{Bn}$ and $M_A$ for a density compression ratio $r=\rho_2/\rho_1=2$ on the left panel, and $r=2.6$ on the right panel. Values of the anisotropy for which the second law of thermodynamics is violated are not plotted. The surfaces indicates shock parameters for which the firehose instability can become unstable.}
  \end{figure*}

\begin{figure*}
\includegraphics[height=8.3cm,width=10.cm, keepaspectratio]{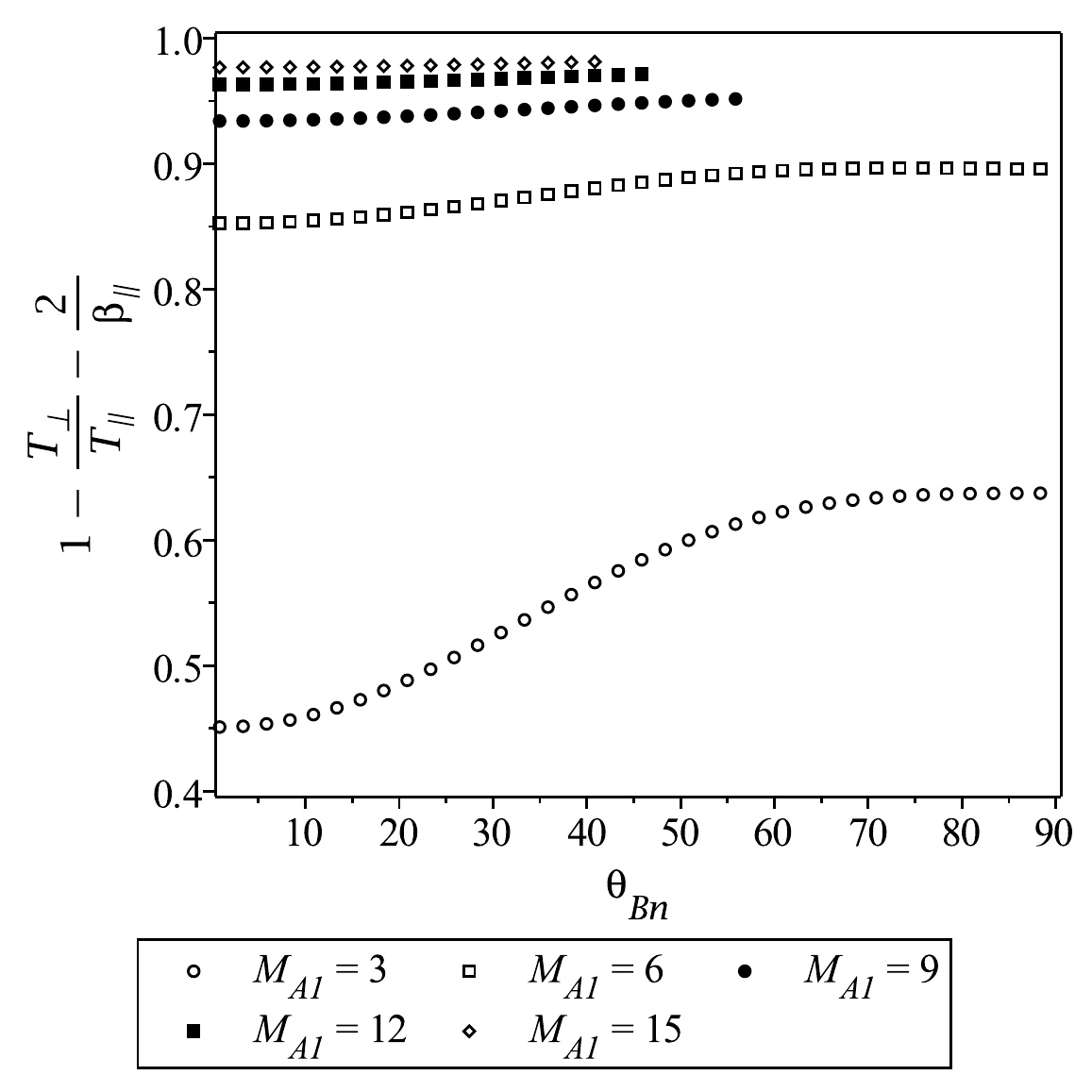}
\includegraphics[height=8.3cm,width=10.cm, keepaspectratio]{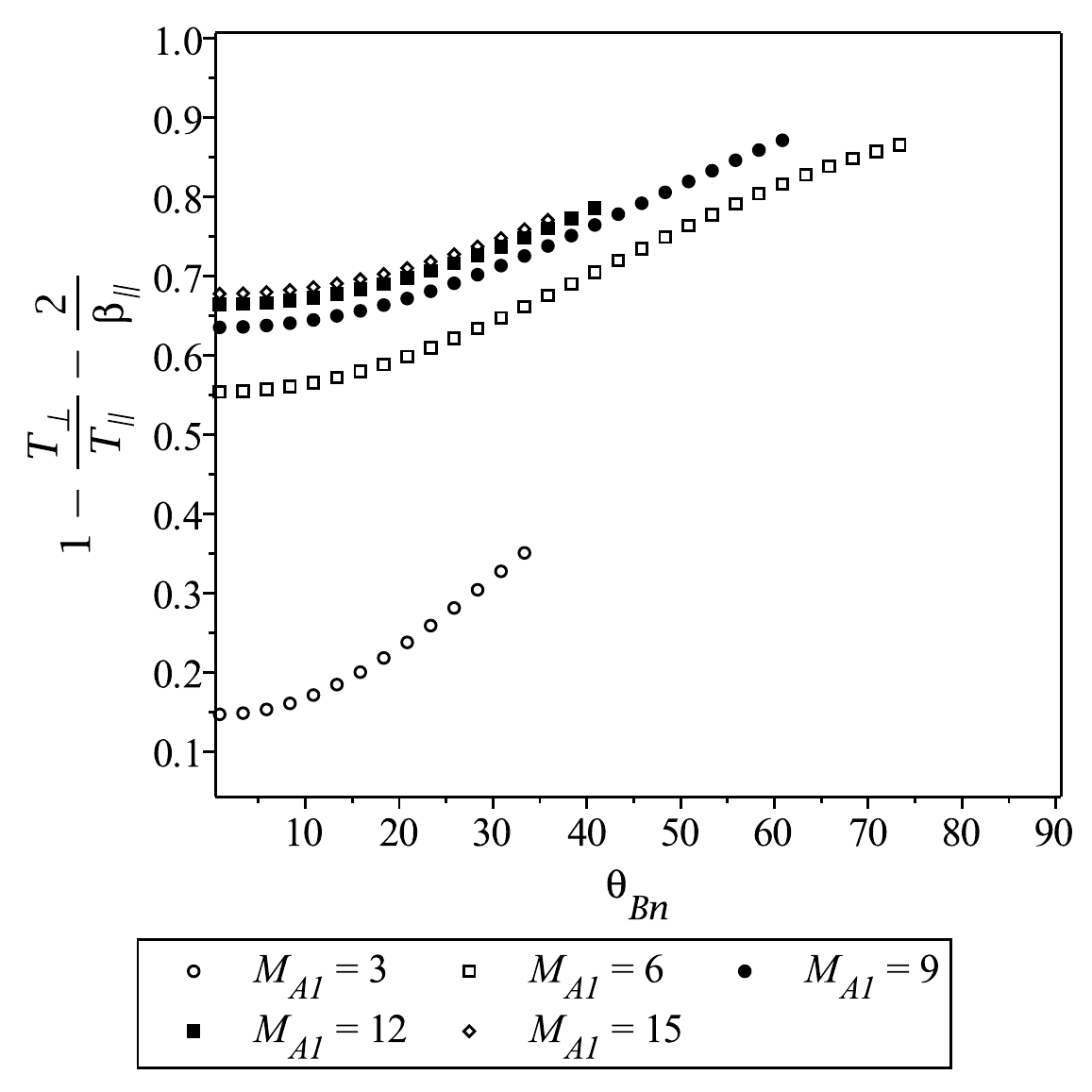}
  \label{fig:sfig2}
  \caption{Firehose instability criteria as a function of $\theta_{Bn}$ and parametrised for Mach numbers values $M_A$ ranging between 3 and 15. The left panel is for a density compression ratio of $r=\rho_2/\rho_1=2$, and the right panel is for $r=2.6$.}
  \end{figure*}

\bibliography{apssamp}

\end{document}